\documentclass[12pt]{article}
\usepackage{arxiv}
\newif\ifclean
\cleantrue  % hide extra markup
\usepackage{graphicx}
\usepackage{amsmath}
\usepackage{amssymb}
\usepackage{amsfonts}
\usepackage{subcaption}
\graphicspath{{{./}}}

\ifclean
\newcommand{\COMMENT}[1]{}
\newcommand{\QUESTION}[1]{}
\else
\usepackage{color}
\newcommand{\COMMENT}[1]{\textcolor{cyan}{{[ \sc{#1} ]}}} % comments
\newcommand{\QUESTION}[1]{\textcolor{cyan}{{[ Q: \sc{#1} ]}}} % questions
\newcommand{\red}[1]{\textcolor{red}{{#1}}}
\newcommand{\caution}{\red{\bf Draft: \today. Do not distribute.}}
\fi

\usepackage[color=yellow]{todonotes}

\usepackage{amsthm}

\newcommand{\fref}[1]{Fig.\,\ref{#1}}
\newcommand{\tref}[1]{Table\,\ref{#1}}
\newcommand{\eref}[1]{Eq.\,(\ref{#1})}

\newcommand{\sref}[1]{Sec.\!~\ref{#1}}

\newcommand{\cref}[1]{Ref.\,\cite{#1}}

\newcommand{\ie}{{\it i.e.}\!\, }
\newcommand{\eg}{{\it e.g.}\!\, }

\newcommand{\etal}{{\it et al.} }

\newcommand{\Cbb}{\mathbb{C}}

\newcommand{\cb}{\mathbf{c}}

\newcommand{\fb}{\mathbf{f}}

\newcommand{\gb}{\mathbf{g}}

\newcommand{\rb}{\mathbf{r}}
\newcommand{\xb}{\mathbf{x}}

\newcommand{\Ab}{\mathbf{A}}
\newcommand{\Bb}{\mathbf{B}}
\newcommand{\Cb}{\mathbf{C}}

\newcommand{\Eb}{\mathbf{E}}
\newcommand{\Fb}{\mathbf{F}}

\newcommand{\Qb}{\mathbf{Q}}

\newcommand{\Nb}{\mathbf{N}}

\newcommand{\Ub}{\mathbf{U}}

\newcommand{\Ib}{\mathbf{I}}
\newcommand{\Rb}{\mathbf{R}}
\newcommand{\Sb}{\mathbf{S}}

\newcommand{\Xb}{\mathbf{X}}

\newcommand{\Dc}{\mathcal{D}}

\newcommand{\Ic}{\mathcal{I}}

\newcommand{\Nc}{\mathcal{N}}

\newcommand{\Uc}{\mathcal{U}}

\newcommand{\gs}{\mathsf{g}}

\newcommand{\Xs}{\mathsf{X}}

\newcommand{\Ws}{\mathsf{W}}

\newcommand{\etab}{{\boldsymbol{\eta}}}

\newcommand{\phib}{{\boldsymbol{\phi}}}

\newcommand{\Omegab}{{\boldsymbol{\Omega}}}

\newcommand{\tr}{{\operatorname{tr}}}

\newcommand{\dev}{{\operatorname{dev}}}

\newcommand{\partialb}{{{\boldsymbol{\partial}}}}
\newcommand{\NN}{{\mathsf{N}\!\mathsf{N}}}

\newcommand{\basis}{{\mathcal{B}}}
\newcommand{\invariants}{{\mathcal{I}}}

\newcommand{\dd}{{\mathrm{d}}}
\newcommand{\consistencyrate}{{\gamma}}
\newcommand{\dissipation}{{\Gamma}}
\newcommand{\Dt}{{\Delta\! t}}

\newcommand{\relu}{{\operatorname{ReLU}}}
\newcommand{\hardsigmoid}{{\operatorname{HardSigmoid}}}
\newcommand{\softplus}{{\operatorname{SoftPlus}}}
\newcommand{\ICNN}{{\mathsf{ICNN}}}
\newcommand{\defgrad}{\mathbf{F}}
\newcommand{\stress}{\mathbf{S}}
\newcommand{\cauchy}{\boldsymbol{\sigma}}
\newcommand{\strain}{\mathbf{E}}

\newcommand{\energy}{\Psi}
\newcommand{\internalstate}{\mathbf{h}}
\newcommand{\conjugateforce}{\mathbf{K}}
\newcommand{\yield}{\Upsilon}
\newcommand{\rhs}{\rb}
\newcommand{\features}{\phib}

\renewcommand{\shorttitle}{\it Attention-based neural ordinary differential equation framework}

\title{\bf\Large An attention-based neural ordinary differential equation framework for modeling inelastic processes}
\author{
Reese E. Jones\textsuperscript{\footnotemark[1]} \\
Sandia National Laboratories \\
Livermore CA, USA
\And
Jan N. Fuhg\\
The University of Texas at Austin \\
Austin TX, USA
}

\begin{document}

\ifclean
\date{}
\else
\date{\caution}
\fi

\maketitle

\footnote[1]{corresponding: {\tt rjones@sandia.gov}}

\begin{abstract}
To preserve strictly conservative behavior as well as model the variety of dissipative behavior displayed by solid materials, we propose a significant enhancement to the internal state variable-neural ordinary differential equation (ISV-NODE) framework.
In this data-driven, physics-constrained modeling framework internal states are inferred rather than prescribed.
The ISV-NODE consists of: (a) a stress model dependent, on observable deformation and inferred internal state, and (b) a model of the evolution of the internal states.
The enhancements to ISV-NODE proposed in this work are multifold: (a) a partially input convex neural network stress potential provides polyconvexity in terms of observed strain and inferred state, and (b) an internal state flow model uses common latent features to inform novel attention-based gating and drives the flow of internal state only in dissipative regimes.
We demonstrated that this architecture can accurately model dissipative and conservative behavior across an isotropic, isothermal elastic-viscoelastic-elastoplastic spectrum with three exemplars.

\end{abstract}


\COMMENT{internal reviewer: Coleman, Tom, Ravi}

\COMMENT{reviewers: Tepole, Kalina, Bouklas}

\COMMENT{journal: CMAME, IJNME, CM, JApplMech, ZAMM, JMCMC, JCompPhys, }

\section{Introduction}

Solid materials can exhibit a variety of behaviors including elasticity, plasticity, viscosity/creeping, fracture/failure, etc.
Each of the categories of material phenomenology has engendered an abundance of models \cite{truesdell2004non,Malvern1969,Gurtin2010}.
These phenomenological idealizations have a limited range of applicability, yet, subject to atomic configuration and therefore phase changes, materials have the potential to display any of the categorical behaviors in some thermomechanical regime \cite{noll1955continuity,noll1958mathematical,mcbride2018dissipation}.
For example, metals have an elastic region where reversible lattice stretching dominates but at some point, deformation is mediated by irreversible dislocations, while long molecules in a polymer allow dissipative, creeping flow.
Yet, an elastic-plastic metal can become viscous at a high enough temperature or strain rate,  and a soft polymer can become glassy and brittle at a low enough temperature.

It follows that there is a need for a general framework to model constitutive behavior, which is particularly true for \emph{data-driven} models of complex, exotic materials that may be hard to categorize.
Furthermore, this multitude of categories creates a model selection process and an implementation burden.
This need for an encompassing constitutive modeling framework was recognized in the early days of the \emph{Rational Mechanics} movement.
A formulation for a general memory-dependent constitutive was put forward in the pioneering work of Green and Rivlin \cite{green1957mechanicsI,green1959mechanicsII,green1959mechanicsIII} and in the treatise of Truesdell and Noll \cite{truesdell2004non}.
\emph{Inelasticity} as a concept encompassing all material behavior that is dissipative is a more recent development \cite{Simo1998}.
Without exception, each of these models introduces specialized state variables which are largely unobservable.

Contemporary with Green, Rivlin, Truesdell, and Noll,  Coleman-Gurtin \cite{coleman1967thermodynamics} provided a \emph{Internal state variable} (ISV) framework based  on thermodynamic reasoning which was  facilitated by the Coleman-Noll hypothesis
\cite{coleman1972thermodynamics} connecting the generalized forces and displacements that govern internal states.
Later, Halphen and Nguyen \cite{halphen1975materiaux} introduced a similar theory of \emph{Generalized standard material} (GSM) which provides more structure for generalized force conjugacy via a dissipation potential.
These ideas have been extended by Ziegler and Wehrli \cite{ziegler1987derivation} and then further formalized by Houlsby and Puzrin \cite{puzrin2001thermomechanical,houlsby2002rate}.

Our goal is to make progress toward a universal data-driven, physics-constrained inelastic model that can handle viscous effects and sharp elastic-plastic transitions without preconceived/prior assumptions.
Our previous ISV-NODE model \cite{jones2022neural} largely follows ISV theory and is comprised of a stress potential and an internal state flow model.
The internal states are inferred from data.
In this work, we achieve a significant improvement of the original ISV by enhancing the flow model with a novel gating/attention \cite{vaswani2017attention} mechanism that provides sharp conservative/dissipation transitions when necessary/appropriate.
In addition, we use a partially input convex neural network (ICNN) \cite{amos2017input} for the free energy to provide polyconvexity in the observable strain invariants and no restrictions to the inferred internal state dependencies.

In the next section, \sref{sec:related}, we put our contribution in the context of recent related work using neural networks (NNs) for constitutive modeling.
Then, in \sref{sec:theory}, we survey the universal principles that guide constitutive model construction.
\sref{sec:model} provides details of the proposed ISV-NODE framework.
Then, in \sref{sec:results}, we provide demonstrations of the proposed model in hyperelastic, finite viscoelastic, and elastoplastic regimes.
Lastly, \sref{sec:conclusion} provides a discussion of results and avenues for future work.

\section{Related work} \label{sec:related}

Data-driven, physics-constrained neural network models of inelastic material behavior,
where the assumption of an internal state is necessary, is an active field of research \cite{chen2021recurrent,
xu2021learning,
as2023mechanics,
wang2023automated,
rosenkranz2024viscoelasticty,
abdolazizi2024viscoelastic,
thakur2024viscoelasticnet,
qin2024physics,
mozaffar2019deep,
koeppe2019efficient,
gorji2020potential,
wu2020recurrent,
abueidda2021deep,
bonatti2022importance,
borkowski2022recurrent,
rosenkranz2023comparative,
heidenreich2024transfer,
heidenreich2024recurrent,
boes2024accounting,
tac2024data}.
For more information, we refer to our recent review on the topic \cite{fuhg2024review}.
For brevity, we will only discuss selected works to put the current developments in context.

Now a foundation for constitutive modeling with NNs akin to hyperelasticity in finite deformation theory, the use of input convex NNs \cite{amos2017input} to impose polyconvexity on a strain energy potential has seen widespread application \cite{tac2022data,chen2022polyconvex,as2022mechanics,xu2021learning,klein2022polyconvex,klein2023parametrized,kalina2024neural,fuhg2022learning,fuhg2022machine}.
Partially convex ICNNs \cite{amos2017input} and monotonic neural networks \cite{fuhg2023modular} have seen fewer applications in constitutive modeling \cite{jadoon2024inverse, jadoon2025automated,jadoon2025input} but are necessary to handle the inferred nature of data-driven internal states in this work.

The history dependence of inelastic materials has been modeled in various ways with NNs.
The seminal work of Ghaboussi and Sidarta \cite{ghaboussi1998new} introduced the use of NNs to model elastoplastic response using an incremental stress formulation akin to hypoelasticity \cite{truesdell2004non}, which resembles a recurrent neural network (RNN) \cite{hochreiter1997long,cho2014learning}.
Later, Xu \etal \cite{xu2020inverse} developed another hypoelasticity-like NN model of viscoelastic behavior and trained it via optimization constrained by boundary value problem solutions.

RNNs take into account the causality of time-series data and have been widely employed to model
viscoelasticity
\cite{chen2021recurrent,
xu2021learning,
wang2023automated,
rosenkranz2024viscoelasticty,
abdolazizi2024viscoelastic,
qin2024physics}
and elastoplasticity
\cite{mozaffar2019deep,
gorji2020potential,
wu2020recurrent,
borkowski2022recurrent,
heidenreich2024transfer,
heidenreich2024recurrent,
ghane2024recurrent}.
They also have inherent gating mechanisms that allow them to focus attention on long and short-term behavior \cite{hochreiter1997long,cho2014learning}.
This gating mechanism has similarities with the inspiration for this work,  the highly cited attention paper in the machine learning literature by Vaswani \etal \cite{vaswani2017attention}.
Bonatti and Mohr \cite{bonatti2021one} developed a parsimonious  \emph{Minimal State Cell} type of RNN in pursuit of a universal material model albeit in an incremental hypoelastic-like form.
Bonatti \etal \cite{bonatti2022importance} demonstrated the effectiveness of a particular type of space-filling trajectories in training RNN material models;  we use simple space-filling random walks to generate the training data in this work.

Crystal plasticity has been a particular focus of NN research since homogenized models can greatly accelerate mesoscale simulations.
Mozaffar \etal \cite{mozaffar2019deep} demonstrated the effectiveness of RNNs in modeling the homogenized response of polycrystals.
Contemporary with Mozaffar \etal \cite{mozaffar2019deep},  Frankel \etal \cite{frankel2019predicting} used a convolutional neural network (CNN) and a RNN to predict the mechanical response of oligocrystals.
This was followed by Frankel \etal \cite{frankel2020prediction} which adopted a combined CNN-RNN to predict the evolving stress fields of polycrystals.
Later Bonatti \cite{bonatti2022cp} compared an RNN model to the more traditional use of fast Fourier transforms in simulating crystal plasticity.

Although less widely adopted in inelastic modeling, neural ordinary differential equations (NODEs) \cite{chen2018neural,dupont2019augmented} where an ordinary differential equation is driven by a NN right-hand side have a number of advantages: the stability and accuracy theory of more familiar classical physical simulation is directly applicable, and they enable a direct transposition of thermodynamic and constitutive theory to a learning framework.
Tac \etal  \cite{tac2022dataA,tac2022dataB,tacc2023data} have been particularly active in the application of NODEs to viscoelasticity with a focus on soft/tissue mechanical response.
Contemporary with Tac \etal, Jones \etal \cite{jones2022neural} adapted an augmented NODE \cite{dupont2019augmented} to model homogeneous viscoelastic and elastic-plastic materials as well as stochastic volumes of composites.
Within the Coleman-Gurtin ISV framework \cite{coleman1967general}, Jones \etal \cite{jones2022neural}
proposed to infer, not prescribe, the internal variables and demonstrated this as an effective approach with potential and tensor-basis formulation \cite{ling2016machine}.
Hauth \etal \cite{hauth2024uncertainty} demonstrated a novel uncertainty quantification technique on a NODE material model.
Following these works, Rodrigues \etal  \cite{rodrigues2024finding} used the related \emph{universal differential equations} (UDE) \cite{rackauckas2020universal} learning architecture to model viscoelasticity.
Also related is the work of As'ad and Farhat \cite{as2022mechanics} which provided an alternative, integral-based, time history/memory kernel approach specific to modeling viscoelasticity.

More recently a number of developments \cite{rosenkranz2023comparative,rosenkranz2024viscoelasticty,flaschel5023581convex} have been based on the GSM framework of Halphen and Nguyen \cite{halphen1975materiaux}.
Rosenkranz \etal \cite{rosenkranz2023comparative} provides a comparative study of different formulations of GSM with feedforward NNs (FFNNs) and RNNs, while Rosenkranz \etal \cite{rosenkranz2024viscoelasticty} used a partially convex ICNN and an RNN to model viscoelasticity.
Flaschel \etal \cite{flaschel5023581convex} employed the framework to model elastic, viscoelastic, plastic, and viscoelastic behavior for simplified loading such as biaxial tension.

\section{Constitutive principles} \label{sec:theory}

In this section, we briefly outline principles that constrain a phenomenological material model to be well-behaved in elastic and inelastic regimes.
We frame the exposition in terms of the strain energy density $\energy$ of a homogenous, isotropic, isothermal material for the present developments.
A more in-depth discussion can be found in the classic texts \cite{truesdell2004non,Malvern1969,Simo1998,Gurtin2010,eringen2013continuum}.

\subsection{Coordinate frame equivariance}
The fact that a model should not depend on a particular frame of reference/coordinate system is a primary principle, which is oftentimes called \emph{material frame indifference} in the mechanics literature.
In particular, an energy density $\energy$ that depends on deformation through the deformation gradient $\defgrad = \partialb_\Xb \hat{\xb}(\Xb,t)$ should be invariant:
\begin{equation}
\energy(\defgrad) = \energy(\Qb \defgrad)
\end{equation}
to a change of coordinates $\xb^\dagger = \Qb \xb + \cb$ with $\Qb \in \text{Orth}$ of the current configuration, where $\xb^\dagger$ is a rigid transformation of $\xb$.
Here $\xb$ is the current placement of a material point identified with position  $\Xb$ in the reference configuration.
If $\defgrad$ is replaced with an objective strain measure \cite{silhavy2013mechanics} which depends only on local stretch $\Ub= \Ub^T = \Ub^\dagger$ then the dependence on the local rotation $\Rb^\dagger = \Qb \Rb$ (from the polar decomposition $\defgrad = \Rb \Ub$ with $\Rb \in \text{Orth}^+$) is eliminated.
The right Cauchy–Green deformation tensor $\Cb \equiv \defgrad^T \defgrad = \Ub^2$ is a common choice as it does not require the polar decomposition of deformation.
Note this is one choice in a family of strain measures \cite{seth1964generalized,hill1968constitutiveI,hill1968constitutiveII,doyle1956nonlinear} that may decrease or increase the complexity of the learning/model calibration goal \cite{fuhg2024stress}.

Furthermore, we  assume isotropic material behavior so that the formulation of the strain energy can be reduced to dependence on three scalar deformation invariants:
\begin{equation}  \label{eq:invariant_potential}
\energy(\defgrad) = \hat{\energy}(I_1,I_2,I_3) \ ,
\end{equation}
where, again, we have a choice in what invariants to use \cite{fuhg2024stress}.
The Cayley-Hamilton invariants
\begin{equation} \label{eq:ch_invariants}
I_1 = \tr \Cb, \quad
I_2 = \tr \Cb^* , \quad % = \tfrac{1}{2} ( \tr^2 \Cb - \tr \Cb^2) , \quad
I_3 = \det \Cb
\end{equation}
are a common choice, where $\Cb^* = \det(\Cb) \Cb^{-T}$ is the adjugate of $\Cb$, $I_2 = \tr \Cb^* = \tfrac{1}{2} ( \tr^2 \Cb - \tr \Cb^2)$,  and $\tr \Cb^{-1} = I_2 / I_3$.
These invariants can be related to the deformation of lines, areas, and volumes and proportional to permutationally invariant sums of principle stresses to the first, second, and third powers, respectively.

\subsection{Well-behaved strain energy}

A multitude of theoretic constraints and assumptions \cite{truesdell2004non,silhavy2013mechanics} have been postulated for a well-behaved strain energy potential.
Ball's solvability postulate \emph{polyconvexity} \cite{ball1976convexity} is the most well-accepted condition, where the strain energy $\hat{\energy}$
\begin{equation}
\energy(\defgrad) = \check{\energy}(\defgrad, \defgrad^*, \det \defgrad)
\end{equation}
is required to be convex in each of its arguments, namely the deformation gradient $\defgrad$, its adjugate $\defgrad^* \equiv \det(\defgrad) \defgrad^{-T}$, and its determinant $\det \defgrad$, which directly govern the deformation of line segments, areas, and volumes, respectively.
Applying the invariance principle,  it is sufficient for $\hat{\energy}(I_1,I_2,I_3)$, to be (a) convex and monotonically increasing in $I_{1}$ and $I_{2}$, and (b) convex in  $J = \sqrt{I_3}$,  and, hence, convex and non-decreasing in $I_{3}$ \cite{schroder2003invariant}.
Therefore we reformulate the strain energy in \eref{eq:invariant_potential} as
\begin{equation} \label{eq:elastic_potential}
\energy = \hat{\energy}(I_1,I_2,J) \ ,
\end{equation}
and denote $\invariants_\Cb = \{ I_1,I_2,J \}$

\subsection{Thermodynamics}

As a statement of the second law of thermodynamics, the Clausius-Duhem inequality \cite{silhavy2013mechanics} reduces to
\begin{equation} \label{eq:2nd_law}
\dissipation \equiv  \dot{\energy} - \stress:\dot{\strain} \geq 0
\end{equation}
for a isothermal, purely mechanical process, where the strain energy $\energy$ is identified with the Helmholtz free energy at constant temperature, $\stress$ with the second Piola-Kirchhoff stress, and $\Eb = \tfrac{1}{2} ( \Cb - \Ib )$ with the Lagrange strain.
Nondissipation $\dissipation = 0$ leads to
\begin{equation} \label{eq:stress_potential}
(\partialb_\strain {\energy} - \stress):\dot{\strain} = 0
\quad \Rightarrow \quad
\stress = \partialb_\strain {\energy}  = 2 \partialb_\Cb {\energy}
\end{equation}
and conservative (hyperelastic) behavior through the Coleman-Noll procedure \cite{coleman1963thermodynamics}.
Note, as there are a multitude of strain and stretch measures, there is a plurality of objective rates and energy-conjugate stresses \cite{haupt1989application,haupt1996stress}.
Dissipation $\dissipation > 0$, on the other hand, is concomitant with entropy production and (irreversible) internal state change.
It can be characterized by $\internalstate$ which we introduce into the energy
\begin{equation} \label{eq:inelastic_potential}
\energy = \hat{\energy}(\invariants_\Cb, \internalstate) \ ,
\end{equation}
and provides a memory/history dependence.
The Coleman-Gurtin \cite{coleman1967thermodynamics}
\emph{internal state variable} (ISV) theory is a general framework utilizing this construction  and the Clauius-Duhem inequality \eref{eq:2nd_law} in the form:
\begin{equation} \label{eq:coleman-gurtin}
\dissipation \equiv
\stress : \dot{\strain} - \dot{\energy} =
\stress : \dot{\strain} - ( \partialb_\strain \energy : \dot{\strain} + \partial_\internalstate \energy \cdot \dot{\internalstate} ) =
-\partial_\internalstate \energy \cdot \dot{\internalstate}
\ge 0
\end{equation}
It is assumed that \eref{eq:stress_potential} holds in this case; and, like \eref{eq:stress_potential}, a force conjugate to $\internalstate$ can be defined as $\conjugateforce = -\partial_\internalstate \energy$.
The \emph{generalized standard model} (GSM) \cite{halphen1975materiaux,suquet1983continuum,hackl1997generalized,maugin1997thermomechanics,steinmann2021catalogue} provides a similar, albeit extended, general framework.

Now we provide a brief introduction to a few classical material models for purposes of motivating the form of the proposed ISV-NODE model.
Many current finite deformation inelastic models are predicated on a multiplicative decomposition of the deformation gradient \cite{lee1969elastic} :
\begin{equation} \label{eq:FeFp}
\defgrad = \defgrad_e \defgrad_p \ ,
\end{equation}
where $\defgrad$ is an observable quantity and  $\defgrad_e$ and $\defgrad_p$ are related internal variables.
This composition is interpreted as a plastic (irreversible) transformation from reference to intermediate configuration and then elastic deformation from intermediate to current.
This paradigm and its extension to multiple intermediate configurations have been used in models of viscoelasticity, damage, and viscoplasticity \cite{lubarda2004constitutive,schutte2002geometrically,clayton2003multiscale}.

Many elastoplasticity models \cite{simo1988frameworkI,simo1992algorithms} augment \eref{eq:FeFp} with a yield condition
\begin{equation}
\hat{\yield}(\stress,\internalstate) \le 0
\end{equation}
that circumscribes the elastic region $\yield < 0$ with a norm on the stress and a \emph{consistency multiplier} $\gamma < 0$ that modulates the flow of internal variables that allow the region to evolve, \eg
\begin{equation} \label{eq:plastic_flow}
\dot{\Cb}_p^{-1} = \gamma \hat{\Nb}(\Sb)
\end{equation}
from \cref{Simo1998}, where $\hat{\Nb}$ is the flow direction.
If the right-hand side of \eref{eq:plastic_flow} is derived from a positively homogeneous function of degree one then the flow is rate-independent, otherwise rate effects can be introduced \cite{mcbride2018dissipation,ottosen2005mechanics,kirchheim2016rank}.
In the case of rate-independent plasticity the multiplier $\gamma$ is determined by making the plastic flow consistent with the yield surface and the Karush-Kuhn-Tucker (KKT) optimality conditions:
\begin{equation}
\yield \le 0,\quad \consistencyrate \ge 0,\quad \yield \consistencyrate = 0
\end{equation}
By way of the Coleman-Noll procedure and \eref{eq:coleman-gurtin} the stress is given by \eref{eq:stress_potential}.
A formulation of viscoelasticity \cite{reese1998theory} resembles these plasticity models in that a multiplicative decomposition
\begin{equation}
\defgrad = \defgrad_e \defgrad_i
\end{equation}
is used but the flow is unconstrained by a yield condition (or, more generally, a degree one homogeneity requirement on a potential resembling an L1 norm \cite{mcbride2018dissipation}).
In this particular model a decomposition of energy similar to a Maxwell model
\begin{equation}
\energy = \energy_\text{eq}(\Cb) + \energy_\text{neq}(\Cb_e)
\end{equation}
is used.
The condition \eqref{eq:coleman-gurtin}  leads to
\begin{equation}
\left(\stress - 2 \partialb_\Cb \energy_\text{eq} - 2 \defgrad_i^{-1} \partialb_{\Cb_e} \energy_\text{neq}  \defgrad_i^{-T} \right) : \dot{\Cb}
- 2 \energy_\text{neq} \dot{\Cb}_e
\ge 0
\end{equation}
and thereby a definition for stress $\stress$.

In these models, the construction of the mechanisms that govern the flow of the internal states is one of the more challenging aspects.
Oftentimes a quadratic product between conjugate forces and flow is assumed to ensure dissipation, which is effectively a dissipation potential.
Other authors explicitly construct a convex dissipation potential $\Phi$ such that
\begin{equation} \label{eq:dissipation_potential}
\dot{\internalstate} =  \partialb_\conjugateforce     \Phi(\invariants,\internalstate)
\end{equation}
as in the GSM theory.

\section{Neural network model} \label{sec:model}

As in \cref{jones2022neural} we seek to infer, rather than prescribe, the variables characterizing the unobservable internal state by designing a model that learns the stress response and its dependence on measurable strain and inferred state inputs while obeying guiding physical principles.
We assume the hidden, internal state $\internalstate$ is composed of scalar invariants (as they are derived from observable invariants through the training process).

An ISV-NODE model consists of
(a) the (observable) stress response $\stress$ derived from a potential $\energy$
\begin{equation} \label{eq:model_response}
\stress = 2 \partialb_\Cb \energy(\Cb,\internalstate)
= 2  \partialb_\Cb \NN_\energy(\invariants_\Cb,\internalstate)
\end{equation}
which is approximated by $\NN_\energy$  that is dependent on the current stretch $\Cb$ and a set of internal state variables $\internalstate$,
and
(b) the evolution of internal state $\internalstate$ governed by the ordinary differential equation
\begin{equation} \label{eq:model_flow}
\dot{\internalstate} = \rhs(\Cb,\dot{\Cb},\internalstate)
=   \NN_\rhs(\invariants_{\Cb,\dot{\Cb}},\internalstate)
\end{equation}
which is driven by a right-hand side approximated by $\NN_\rhs$.
This framework allows us to model inelastic materials as a sequence of elastic materials indexed by $\internalstate$.
The potential $\NN_\energy$ depends only on scalar invariants $\invariants_\Cb$ of the stretch $\Cb$ and internal state $\internalstate$ while flow depends on the joint invariants of stretch and its rate $\dot{\Cb}$, $\invariants_{\Cb,\dot{\Cb}}$, as well as $\internalstate$, to accommodate rate effects.

\begin{figure}
\centering
\begin{subfigure}[c]{0.55\linewidth}
% ././Figures/isv_node.pdf ==> fig1a.pdf
\includegraphics[width=0.95\linewidth]{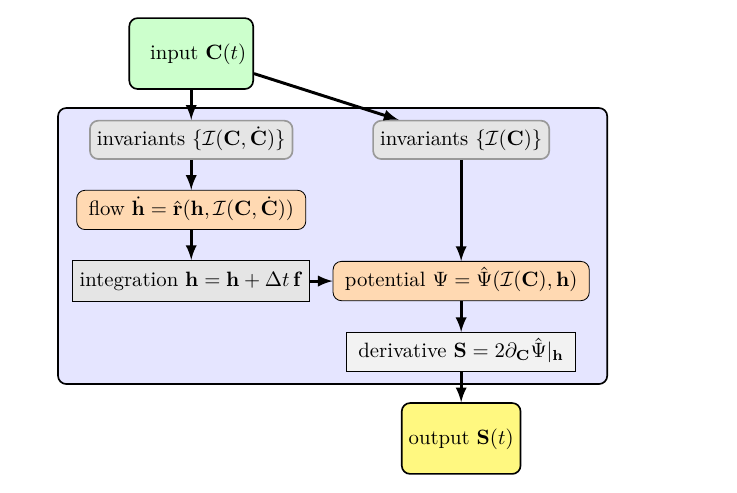}
\caption{ISV NODE}
\end{subfigure}
\begin{subfigure}[c]{0.49\linewidth}
% ././Figures/pot.pdf ==> fig1b.pdf
\includegraphics[width=0.95\linewidth]{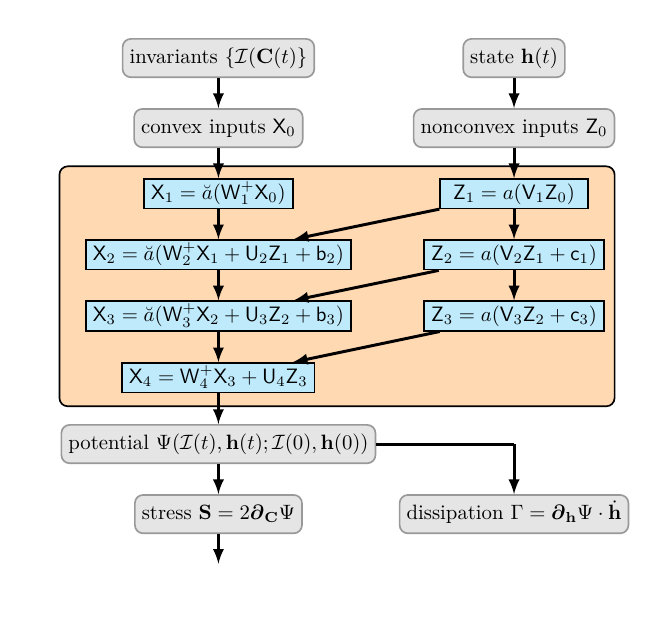}
\caption{Stress potential}
\end{subfigure}
\begin{subfigure}[c]{0.49\linewidth}
% ././Figures/rhs.pdf ==> fig1c.pdf
\includegraphics[width=0.95\linewidth]{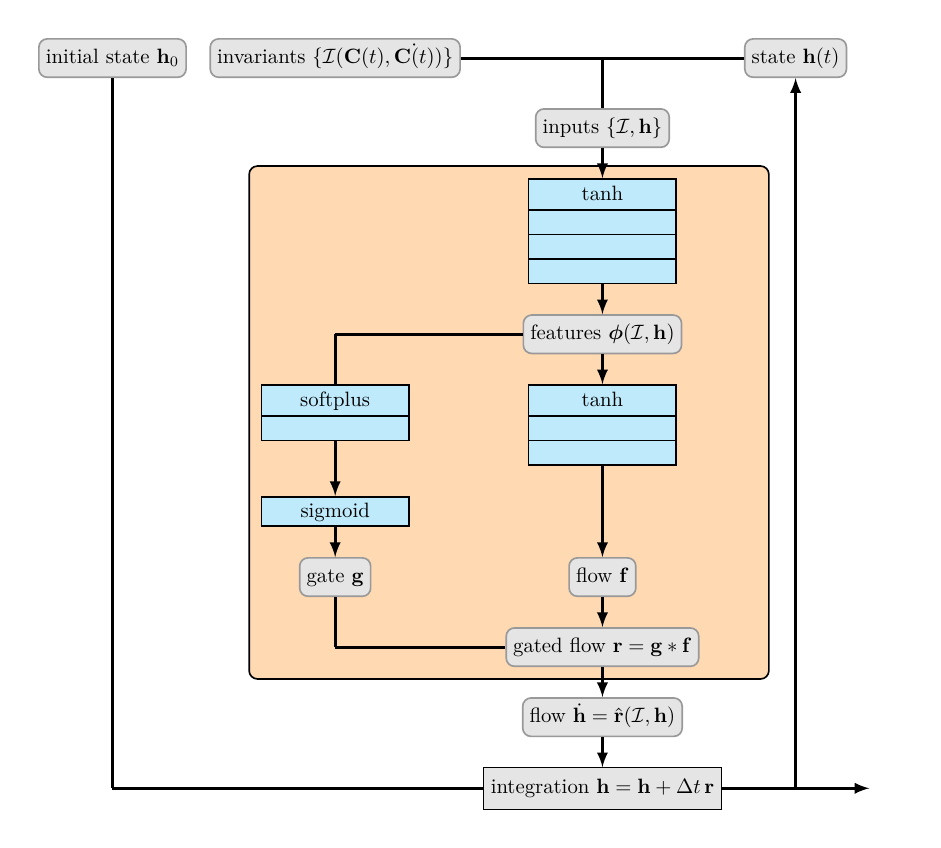}
\caption{State flow}
\end{subfigure}
\caption{Model architecture: (a) ISV-NODE, (b) stress potential component, (c) internal state flow component.
For (b) and (c) the orange boxes correspond to those in (a) and the trainable aspects are in blue.
}
\label{fig:architecture}
\end{figure}

\fref{fig:architecture}a illustrates the connections between these two components in transforming the input stretch $\Cb(t)$ to the output stress $\stress(t)$.
First the invariants $\invariants_{\Cb}$ and $\invariants_{\Cb,\dot{\Cb}}$ are extracted from $\Cb(t)$.
The stretch and rate invariants are fed to the flow integrator.
The resulting hidden state $\internalstate$ is combined with the stretch invariants to evaluate the derivative of the potential with respect to stretch $\Cb$, which becomes the output stress $\stress$.

\subsection{Potential}
We use a partially convex NN to form the free energy
\begin{equation}\label{eq:picnn_pot}
\energy = \ICNN(\invariants_\Cb, \internalstate)
\end{equation}
in terms of the observable stretch invariants $\invariants_\Cb$ and inferred internal state $\internalstate$.
The stretch invariants  $\invariants_\Cb = {I_1, I_2, J}$ have a corresponding (derivative) basis
\begin{equation}
\basis =
\{ \partialb_\Cb I_a \}_{a=1}^3
= \{ \Ib, \tr(\Cb) \Ib - \Cb, \tfrac{1}{2} J \Cb^{-1} \}
\end{equation}
So the stress is given by
\begin{equation} \label{eq:stress}
\Sb \equiv 2 \partial_\Cb \energy
= 2 \sum_a \partial_{I_a} \energy  \, \Bb_a
= 2 \sum_a \partial_{I_a} \hat{\energy} \partialb_{\Cb} I_a
= 2 \left(
\partial_{I_1} \hat{\energy} \Ib +
\partial_{I_2} \hat{\energy} (I_1 \Ib - \Cb) +
\partial_{J  } \hat{\energy} (\frac{1}{2} J \Cb^{-1})
\right)
\end{equation}

We enforce two principles on the form of the NN for the potential: (a) polyconvexity (refer to \sref{sec:theory}), and (b) a stress-free reference response (to ease connecting with experimental data).

\paragraph{(a) Polyconvexity}
Embedding polyconvexity in a hyperelastic stress potential has been done by many authors \cite{tac2022data,chen2022polyconvex,as2022mechanics,xu2021learning,klein2022polyconvex,klein2023parametrized,kalina2024neural,fuhg2022learning,fuhg2022machine}.
Following \cite{amos2017input,fuhg2024polyconvex}, we employ a partially \emph{input convex neural network} (ICNN) to represent the stress potential,
\begin{equation}
\energy(\invariants_\Cb, \internalstate) = \ICNN(\invariants_\Cb, \internalstate)
\end{equation}
This ICNN is only convex and monotonically increasing in its first argument $\invariants_\Cb$ to ensure polyconvexity with respect to the observable stretch invariants \cite{ball1976convexity}
Note derivatives of this potential with respect to $\Cb$ and $\internalstate$ provide the stress $\stress$ \eqref{eq:stress} and the internal state conjugate forces $\conjugateforce$, respectively.
The latter is used in measuring the dissipation $\dissipation$ \eqref{eq:dissipation}.

The structure of this particular adaptation of the ICNN is diagrammed in \fref{fig:architecture}b and consists of two stacks of layers such that the output of the nonconvex channel feeds into the convex ones.
The weights of the convex layers $\Ws^+_i$ are constrained to be positive and the activations $\breve{a}$ are required to be convex and non-decreasing.
We use $\softplus$ activations for both $\breve{a}$  and $a$ for simplicity.
Note, as in
Linden \etal \cite{linden2023neural}, we also expand invariants to admit negative stresses
$\invariants_\Cb = \{ \tr\Cb, \tr\Cb^*, J, -2 J \}$.

\paragraph{(b) Stress-free reference}

At the reference configuration $\Cb = \Ib$ the NN stress in \eref{eq:stress} is generally not zero:
\begin{equation}
\stress_0  \equiv
\hat{\stress}(\invariants_0, \internalstate_0)
= \underbrace{2 \left(
\partial_{I_1} \tilde{\energy}  +
2 \partial_{I_2} \tilde{\energy}    +
\frac{1}{2} \partial_{J  } \tilde{\energy}
\right)}_{p_0} \Ib \ ,
\end{equation}
where $\Ic_0 = \{ 3, 3, 1 \}$ and $\internalstate_0 = \{0,0,\ldots,0 \}$,
nor is the reference energy
\begin{equation}
\energy_0  = \tilde{\energy}(\invariants_0)
\end{equation}
Following
\cite{kalina2022automated,fuhg2024polyconvex}
with direct extension to the additional internal state dependence, the NN response can be adjusted
\begin{equation}
\energy =  \hat{\energy} - \energy_0 - p_0 (J-1)
\end{equation}
so that the reference state is stress-free and is the datum for energy.
Note $J-1$ has value $0$ and derivative $\Ib$ at the reference state.

\subsection{Flow}
The flow depends on the joint invariants of two symmetric tensors \cite{rivlin1955further,spencer1962isotropic}, namely the stretch $\Cb$ and its rate $\dot{\Cb}$:
\begin{equation}
\{I_i \} =
\{
\tr \Cb, \tr \Cb^2, \tr \Cb^3,
\tr \dot{\Cb}, \tr \dot{\Cb}^2, \tr \dot{\Cb}^3,
\tr \Cb \dot{\Cb}, \tr \Cb \dot{\Cb}^2, \tr \Cb^2 \dot{\Cb},  \tr \Cb^2 \dot{\Cb}^2
\}
\end{equation}
where
$I_i, i=1,2,3$ are purely dependent on stretch, $I_i, i=4,5,6$ are purely dependent on the rate of stretching and the remainder have mixed dependence.
\cref{jones2022neural} proved that $I_i, i=1..6$, \ie omitting the mixed invariants, is sufficient in the case of one tensor is the rate of the other.
So we reduce $\invariants_{\Cb,\dot\Cb}$ to
\begin{equation} \label{eq:flow_invariants}
\invariants_{\Cb,\dot\Cb} = \{
\tr \Cb, \tr \Cb^2, \tr \Cb^3,
\tr \dot{\Cb}, \tr \dot{\Cb}^2, \tr \dot{\Cb}^3 \}
\end{equation}

As diagrammed in \fref{fig:architecture}c, the right-hand side NN in the proposed framework has more structure than that proposed in \cref{jones2022neural}.
The main innovation is the addition of a gating mechanism, akin to \eref{eq:plastic_flow}, that can turn off the flow of internal states $\dot{\internalstate}$ based on the explicit state $\invariants_{\Cb,\dot{\Cb}}$ and the internal state $\internalstate$.
For parameter parsimony, the inputs to the flow are first reduced to a common set of features $\features(\invariants,\internalstate) = \NN_\features(\invariants,\internalstate)$ which drive the separate flow and gate NNs:
\begin{equation} \label{eq:gate}
\NN_\rhs(\Ic_{\Cb,\dot{\Cb}},\internalstate) = \NN_\text{gate}(\features(\Ic_{\Cb,\dot{\Cb}},\internalstate)) * \NN_\text{flow}(\features(\Ic_{\Cb,\dot{\Cb}},\internalstate))
\end{equation}
Each subcomponent NN, $\NN_\features$,$\NN_\text{gate}$, $\NN_\rhs$, is a densely connected feed-forward NN (aka a multilayer perceptron) with particular activations.
For $\NN_\features$ and $\NN_\rhs$ all but the last layers have $\tanh$ activations for NODE stability considerations \cite{drgona2020spectral} and their final activations are linear, while $\NN_\text{gate}$ uses $\softplus$ preliminary activations and a final sigmoid activation, as in many gating/attention applications \cite{gulcehre2018hyperbolic}.

\paragraph{(a) Non-positive dissipation}

The thermodynamic dissipation requirement \eqref{eq:coleman-gurtin} reduces to
\begin{equation} \label{eq:dissipation}
\partialb_\internalstate \energy \cdot \rhs \le 0.
\end{equation}
As in \cref{jones2022neural} we penalize violations in the loss described in \sref{sec:training}, as they represent entropy-producing flow.
This inequality constraint typically is compatible with the model-data fit, as it does not incur a loss of convergence nor accuracy even with relatively large penalties $\approx 10^{-3}$.

\paragraph{(b) Minimum  complexity flow}

The addition of internal state variables $\internalstate$ enlarges the overall state space ( comprised of the observable invariants $\invariants$ and the inferred $\internalstate$) so that flow typically occurs without gating even in conservative regimes.
We believe this is due to the fact there are many more solutions that involve flow than those that have no change in $\internalstate$.
Hence, we need gating to avoid this fictitious dissipation.
In terms of complexity, we want the simplest flow that fits the data.
The gating just described provides an attention mechanism to determine when to turn off the flow, and we use a \emph{hard sigmoid}:
\begin{equation}
\hardsigmoid(x)
= \relu(x)
- \relu(x-1)
=
\left.
\begin{cases}
0 &: x < 0 \\
1 &: x > 1 \\
x &: \text{else} %  x \in [0,1]
\end{cases}
\right\}
\quad
\in [0,1] \,
\end{equation}
to make the transitions sharp enough to represent non-viscous plasticity, for example.
Furthermore, we apply L1 regularization to the weights of the final layer of the gate NN, $\Ws_\text{gate}$ (and no bias)
\begin{equation} \label{eq:gate_layer}
\gs = \hardsigmoid(\Ws_\text{gate} \Xs_n) \in [0,1]
\end{equation}
so that we obtain sparsely non-zero flow.
Further details will be given in \sref{sec:training}.

\subsection{Numerical implementation}
To update the state $\internalstate$ and the output stress $\stress$ over a time step $\Dt$ we use Heun's explicit midpoint integrator
\begin{eqnarray}
\tilde{\internalstate}_{n+1} &=& {\internalstate}_{n} + \Dt \, \hat{\rhs}(\Cb_{n+1}, \Sb_{n}) \\
\tilde{\Sb}_{n+1} &=& \hat{\Sb}(\Cb_{n+1}, \tilde{\internalstate}_{n+1}) \\
{\internalstate}_{n+1} &=& {\internalstate}_{n} + \Dt \,\frac{1}{2} \left( \hat{\rhs}(\Cb_{n}, \Sb_{n}) + \hat{\rhs}(\Cb_{n}, \tilde{\Sb}_{n}) \right)
\end{eqnarray}
which provides a prediction $\tilde{\Sb}$ of stress which is then corrected at the end of the step.
This prediction of stress can be useful in stress-based gating, which will be discussed in \sref{sec:conclusion}.

\section{Data} \label{sec:data}

We generate synthetic training data that is representative of the common material classes: elastic, viscoelastic, elastoplastic.
For each demonstration we generate 1000 random walks of 100 steps each for training data, likewise we use 200 smoother trajectories of 100 steps for validation.
Material parameters were normalized to generate $O(1)$ stress output.
Different types of loading trajectories were selected for training vs. testing data to generate in and out-of-distribution samples.

\subsection{Loading paths}
The training data was generated with random walks through stretch $\Ub_n \equiv \Ub(t_n)$ space:
\begin{equation}
\Ub_{i+1} = \Ub_i + \Delta\Ub  \quad
\text{where} \quad \Delta\Ub \sim \Uc(-\delta, \delta)
\end{equation}
where $\Ub_0 = \Ib$.
The step $\Delta\Ub$ influences the range of rates $| \dot{\Cb} |_\infty \propto | \Delta\Ub |_\infty^2 / \Dt$ sampled in the walk.
Unphysical steps $\Delta\Ub$ are rejected to keep $\Ub$ positive definite.
The deformation gradient is provided $\defgrad(t) = \Ib + \Ub(t)$ by assuming that the local rotation $\Rb = \Ib$ is fixed at the identity.

Likewise, parameterized trajectories
\begin{equation}
\defgrad(t) = \Ib +  \Ab \odot h(\Omegab t )
\end{equation}
where used for validation data, where $h$ is a periodic function, such as $\sin$ or a sawtooth waveform, and $\odot$ is the Hadamard product.
Both the frequencies, $\Omegab \sim \Uc[0,1.5]$,  and the loading modes,  $\Ab \sim \Uc[0,\delta]$, were sampled from uniform distributions.
Note $\Ab$ was constrained to be symmetric and positive definite.
These periodic paths were chosen since they are cyclic (which illustrates dissipation, as will be shown in \sref{sec:results}), relatively smooth, and sample a variety of rates $\dot{\Cb}$.

We evaluated the following data generating models $\{ {\stress}^* \}$ on the path:
\begin{equation}
\stress_n = \stress^*(\Cb_n) + \etab
\end{equation}
and added white noise $\eta_{ij} \sim \varsigma \Nc(0,1)$, with amplitude $\varsigma$, to simulate stochastic noise.
Nominally the amplitude was set to $\varsigma = 10^{-4}$ and $\etab$ was made to be symmetric.
Note errors  reported \sref{sec:results} are with respect to the noiseless data $\Dc = \{\Cb_n, \stress^*(\Cb_n)\}$.

\begin{figure}
\centering
% ././Figures/invariants.png ==> fig2a.png
\includegraphics[width=0.75\linewidth]{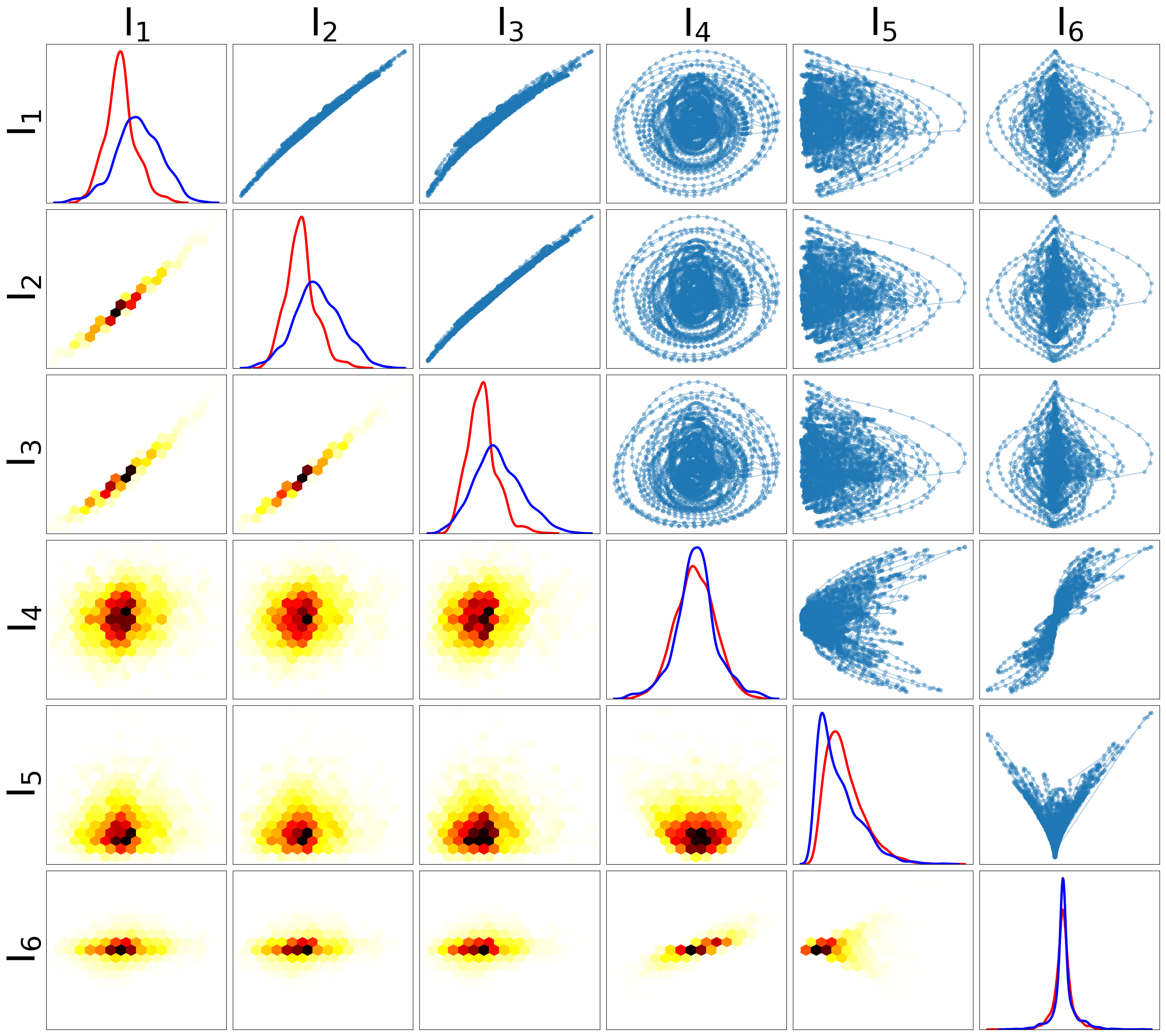}
\caption{Training and testing data invariants.
Upper triangle of panels: testing trajectories,
lower triangle: density of training data
Diagonal panels: comparison of marginalized distributions (red:training, blue:testing) showing in training distribution and out of distribution (testing).
}
\label{fig:invariants}
\end{figure}

\fref{fig:invariants} illustrates the invariants visited by the training data (lower triangle) and validation data (upper triangle).
The (on diagonal) marginal distributions (red: training, blue: validation) best illustrate the fact that the two datasets are similar but cover different regions of invariant space.
In particular, the validation data samples more of the higher stretch invariants $I_1,I_2,I_3\equiv J$ and yet explores similar rates.

\subsection{Data generating models}
For the purpose of demonstrating that the proposed model can span hyperelastic, viscoelastic, and elastoplastic behavior we use three exemplars as surrogates for experiments on these three material response categories with a common set of normalized material parameters.
The hyperelastic response is conservative with no dissipation nor entropy generation.
The viscoelastic response is a dissipative nonequilibrium response with no purely elastic regime.
The elastoplastic response has a circumscribed elastic regime and switches between conservative and dissipative behavior.

\paragraph{Hyperelasticity}

For a hyperelastic exemplar we use a (polyconvex) Rivlin-type \cite{rivlin1997collected} compressible  neo-Hookean model with strain energy
\begin{equation} \label{eq:neohooke}
\energy = \frac{1}{2} \mu \bar{I}_1 + \frac{1}{2} \kappa (J-1)^2 \ ,
\end{equation}
where $\bar{I}_1 = J^{-2/3} \tr \Cb = I_3^{-1/3} I_1  > 0$ and $J = \sqrt{\det \Cb} > 0$.
We take the normalized material properties, $\kappa=1.67$ and $\mu=0.77$, to be representative of a polymer.

The resulting stress measure is the second Piola-Kirchhoff stress
\begin{equation}
\Sb = \mu J^{-2/3} (\Ib - \tfrac{1}{3} I_1 \Cb^{-1})
+ \left( \kappa (1-J^{-1}) \right) \Cb^{-1}
\end{equation}
which corresponds to the Cauchy stress
\begin{equation} \label{eq:cauchy_stress}
\cauchy \equiv \tfrac{1}{J} \defgrad \stress \defgrad^T
= \mu J^{-5/3} \dev(\Bb) + \left(\kappa (J-1) \right) \Ib \ ,
\end{equation}
where $\Bb = \defgrad \defgrad^T$.

\paragraph{Viscoelasticity}
As an alternative to the formulation in \cref{reese1998theory} we use a hereditary integral type viscoelastic model:
\begin{equation}
\cauchy = p_e \Ib + \int_{-\infty}^t G(t-s) \dev \, \dot{\cauchy}_e(\Cb(s)) \dd s
\end{equation}
employing the same neo-Hookean equilibrium model \eqref{eq:neohooke} as the nonlinear kernel.
Here $p_e$ is the pressure associated with
the bulk energy $\tfrac{1}{2} \kappa (J-1)^2$ and $\dev \, \dot{\cauchy}_e(s)$ is the deviatoric part of \eref{eq:cauchy_stress}.
For the isotropic case under consideration, $G$ is a scalar kernel represented by sum of exponentials \cite{puso1998finite,fung2013biomechanics}
\begin{equation}
G(t) = \sum_i G_i \exp(-t/\tau_i)
\end{equation}
for which we employ a three-term Prony series
$\{G_i = 0.8,0.1,0.05 \}$ and $\{ \tau_i = 0.4,0.1,0.01 \}$.

\paragraph{Elastoplasticity}
For the elastoplastic response, we use a finite deformation version of the commonly used J2 model.
We employ a St. Venant model for stress:
\begin{equation}
\Sb = \Cbb : \Eb
= \kappa \, \tr(\Eb) \, \Ib + 2 \mu \, \dev \Eb
\end{equation}
where the bulk modulus $\kappa$ and shear modulus $\mu$ are the same as in the hyperelastic model.
For yield, the usual von Mises-Huber condition:
\begin{equation}
\yield \equiv \| \stress \|_\text{vm} - \Theta(\internalstate) \le 0
\end{equation}
was adopted,  with
\begin{equation}
\| \stress \|_\text{vm} = \tfrac{1}{2} \dev \stress : \dev \stress
\end{equation}
i.e. the $J_2$ invariant of stress and linear hardening
\begin{equation}
\Theta(\epsilon_p) = Y_0 + H \epsilon_p
\end{equation}
with $Y_0 = 0.02 $ and $H = 0.05 $.
The flow is given by an associative rule
\begin{equation}
\dot{\Fb}_p = \dot{\gamma} \partialb_\stress \yield
\end{equation}
with an exponential update \cite{cuitino1992material,Simo1998}.

\section{Training} \label{sec:training}

To calibrate the ISV-NODE and guide an Adam optimizer \cite{kingma2014adam},  we employed a standard mean square error (MSE) loss function
\begin{equation}\label{eq:mse}
\text{MSE} = \frac{1}{N_\text{samples} N_\text{steps}} \sum_{i=1}^{N_\text{samples}} \sum_{j=1}^{N_t}  \| \Sb_i(t_j) - \hat{\Sb}(\Cb_i(t_j),\dot{\Cb}(t_j)) \|^2
\end{equation}
applied to the stress data $\Sb$ and the NN model response $\hat{\Sb}$ \eqref{eq:model_response}.
To enforce the dissipation constraint imposed by the second law \eqref{eq:dissipation} and the complexity penalty \eqref{eq:gate_layer}, we augment the loss $L$ with a penalty
\begin{equation}\label{eq:loss}
L = \text{MSE} + \lambda_\Gamma \langle \Gamma \rangle + \lambda_\text{gate} \| \Ws_\text{gate} \|_1
\end{equation}
where $\varepsilon_\Gamma$ and $\lambda_\text{gate}$ are penalty hyperparameters.
The Macauley bracket of the dissipation $\Gamma$ is implemented as
\begin{equation}\label{eq:pot_penalty}
\langle \Gamma \rangle = \relu(\partialb_\internalstate \Psi \cdot \fb)
\end{equation}
while the gate penalty function is an L1 norm on the weights of the last layer of $\NN_\text{gate}$ which has a $\hardsigmoid$ activation that turns off at zero.

\section{Results} \label{sec:results}

In this section, we demonstrate that the gated ISV-NODE can find $\internalstate$ flows that match the stress data that are dissipative (inelastic viscoelastic and plastic materials) and nondissipative (hyperelastic materials and plastic materials in an elastic state).
The learning and model hyperparameters were not optimized.
We assumed a hidden vector $\internalstate$ of size $N_\internalstate = 9$ was sufficient, while the number of invariants $N_{\invariants_{\Cb}} = 3$ and  $N_{\invariants_{\Cb,\dot{\Cb}}} = 6$ were fixed by the application, as per \sref{sec:model}.
We used a network with 4 layers (including the linear output layer) for the partially convex NN potential $\NN_\energy$, while the flow model had 2 layers for the feature block $\NN_\features$, 2 for the gate block $\NN_\text{gate}$ and 3 for the remaining flow block $\NN_\text{flow}$.
The width of each of these blocks was the input size (potential: $N_{\invariants_{\Cb}}+N_\internalstate$ and flow: $N_{\invariants_{{\Cb},\dot{\Cb}}}+N_\internalstate$ ) plus 8 additional neurons/nodes.
The penalty parameters in \eref{eq:loss} were nominally $\lambda_\dissipation = 10^{-3}$ for the dissipation inequality and $\lambda_\text{gate} = 10^{-4}$ for the complexity/gating penalty.

The error was measured with root mean squared error (RMSE) (\eref{eq:mse}) on normalized data.
In addition, we check power matching
\begin{equation} \label{eq:power_discrepancy}
(\hat{\stress} - \stress) \cdot \dot{\strain} = 0
\end{equation}
as well as the hidden state trajectories and the gating.

\subsection{Hyperelastic}

\begin{figure}
\centering
\begin{subfigure}[b]{0.95\textwidth}
\centering
% ././Figures/elastic_nogate_train-accuracy.pdf ==> fig3a.pdf
\includegraphics[width=0.45\textwidth]{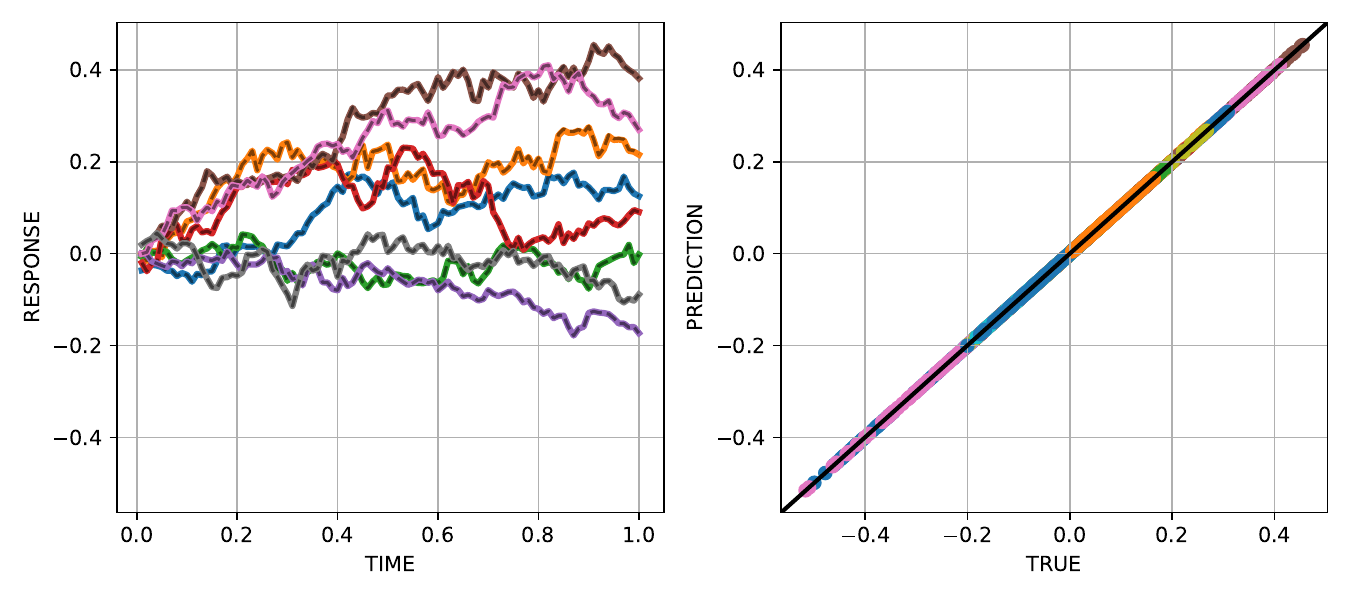}
% ././Figures/elastic_nogate_valid-accuracy.pdf ==> fig3b.pdf
\includegraphics[width=0.45\textwidth]{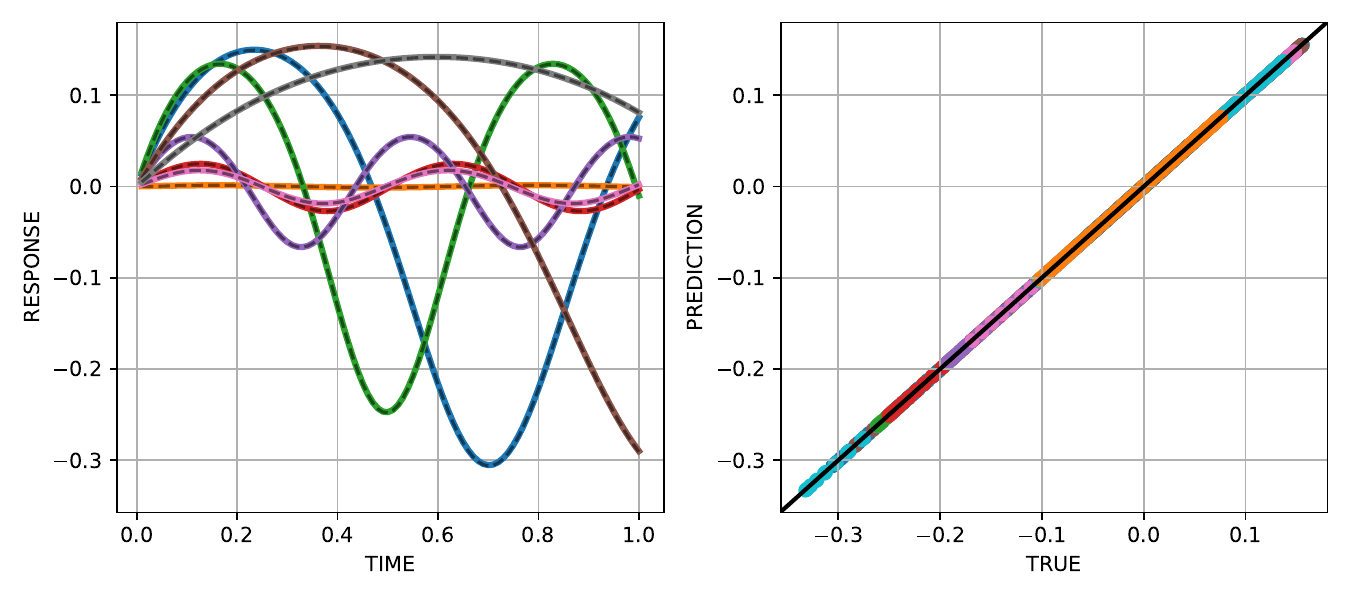}
\caption{no gate}
\end{subfigure}
\begin{subfigure}[b]{0.95\textwidth}
\centering
% ././Figures/elastic_gate_train-accuracy.pdf ==> fig3c.pdf
\includegraphics[width=0.45\textwidth]{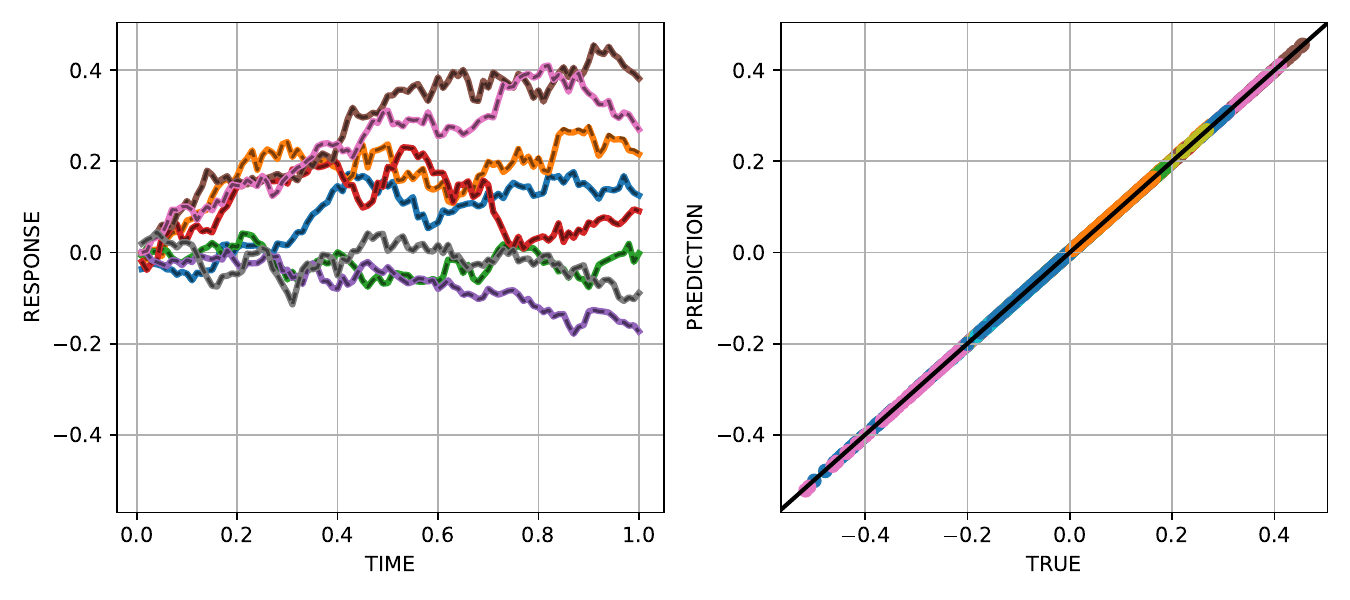}
% ././Figures/elastic_gate_valid-accuracy.pdf ==> fig3d.pdf
\includegraphics[width=0.45\textwidth]{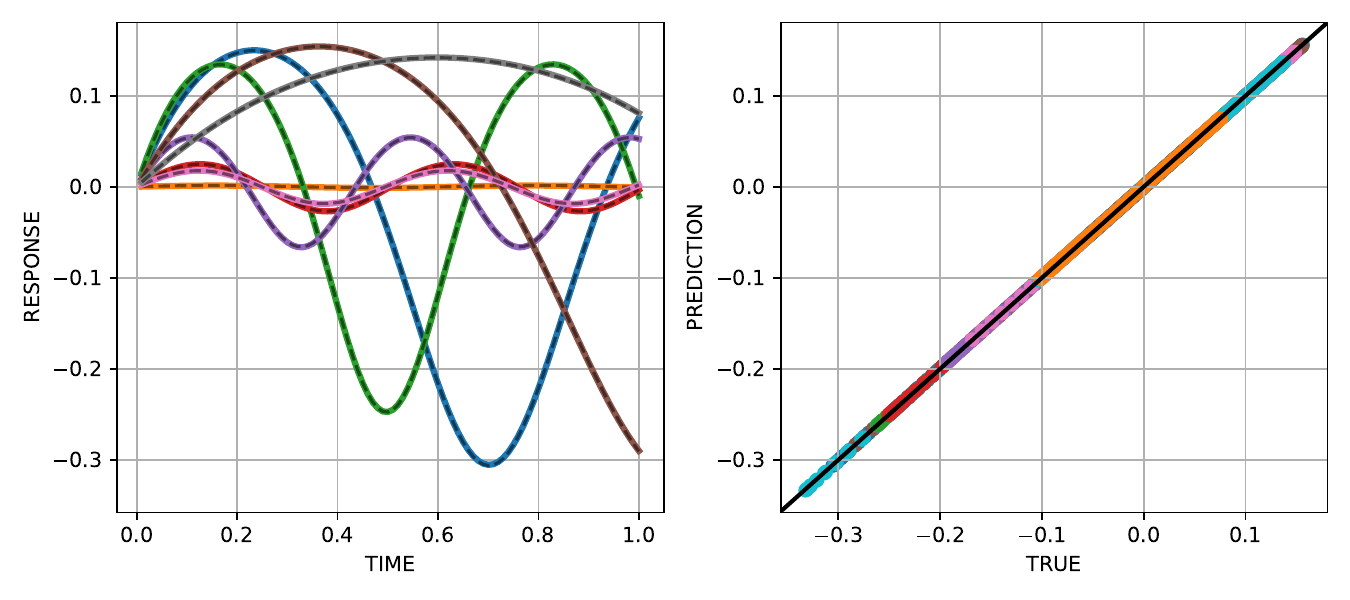}
\caption{gate}
\end{subfigure}
\caption{Hyperelasticity: accuracy for (left) in training distribution, and (right) out of training distribution.
The predictions of stress response versus time (left panels, dashed black) are compared to data (color).
The parity plots (right panels) plot predicted stress samples versus corresponding true values where samples from the same trajectory have the same color. }
\label{fig:elastic_accuracy}
\end{figure}

\fref{fig:elastic_accuracy} compares  8 predicted trajectories to data and the errors for 64 trajectories for both held-out samples of the training distribution (random walks) and samples of the validation distribution (sinusoidal trajectories).
We compare the proposed gated ISV-NODE to a similar architecture with the gate $\NN_\text{gate}$ omitted so that entire flow block is comprised of $\NN_\features$ and $\NN_\text{flow}$ (refer to \fref{fig:architecture}c).
Clearly, the results are nearly indistinguishable, albeit the underlying complexity is different.
\tref{tab:rmse} compares the errors for these two cases (in \fref{fig:elastic_accuracy} ``gate"  is ``high gate" in the table), as well as two more.
The gated model does score slightly better in quantitative error than its ungated counterpart.

\begin{figure}
\begin{subfigure}[b]{0.95\textwidth}
\centering
% ././Figures/elastic_nogate_valid-evolution.pdf ==> fig4a.pdf
\includegraphics[width=0.95\textwidth]{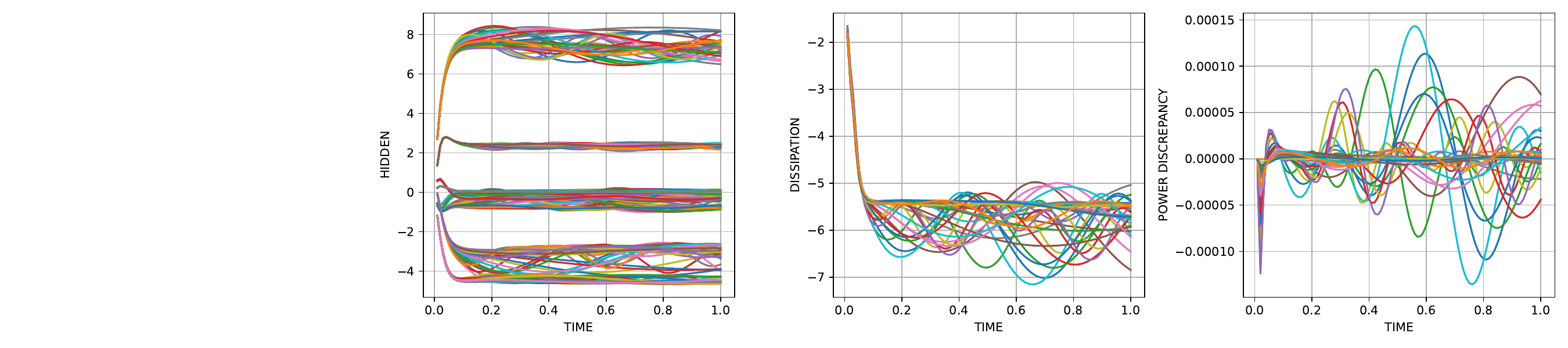}
\caption{no gate}
\end{subfigure}
\begin{subfigure}[b]{0.95\textwidth}
\centering
% ././Figures/elastic_logate_test-evolution.pdf ==> fig4b.pdf
\includegraphics[width=0.95\textwidth]{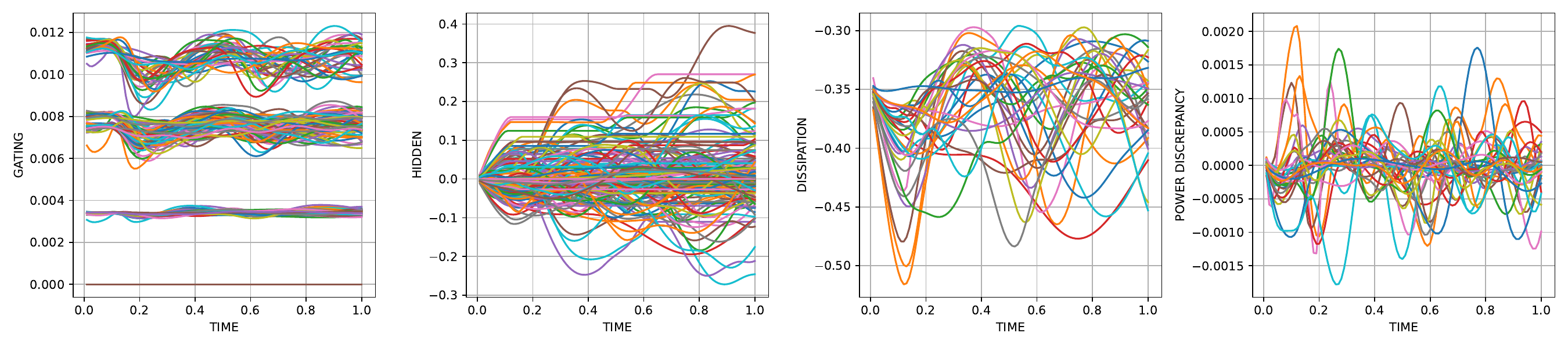}
\caption{low gate}
\end{subfigure}
\begin{subfigure}[b]{0.95\textwidth}
\centering
% ././Figures/elastic_gate_valid-evolution.pdf ==> fig4c.pdf
\includegraphics[width=0.95\textwidth]{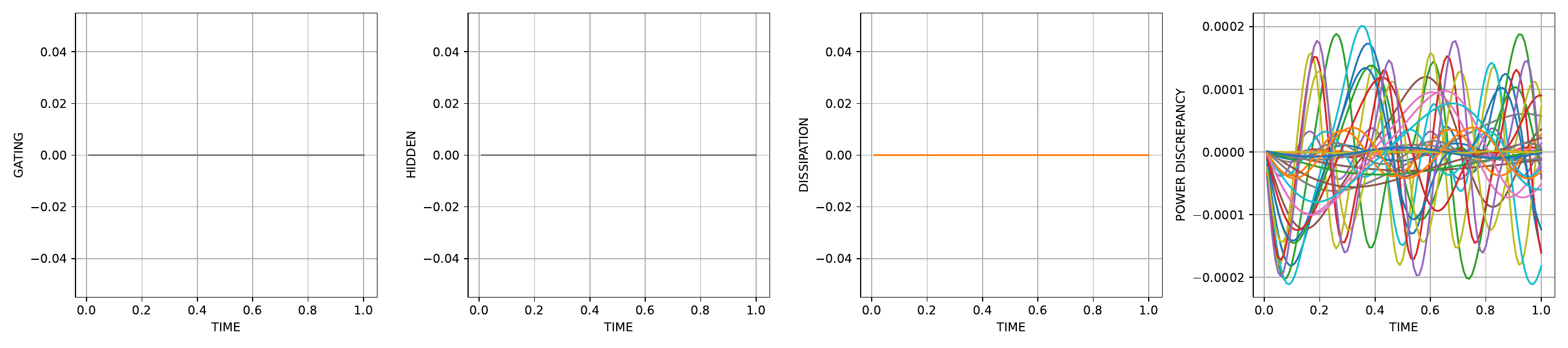}
\caption{high gate}
\end{subfigure}
\begin{subfigure}[b]{0.95\textwidth}
\centering
% ././Figures/elastic_noise_valid-evolution.pdf ==> fig4d.pdf
\includegraphics[width=0.95\textwidth]{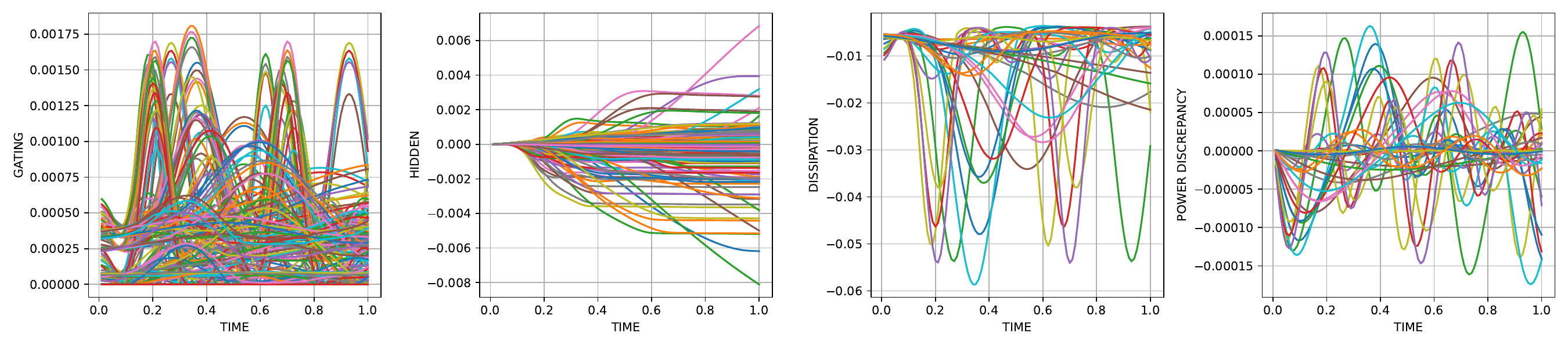}
\caption{high gate with 1\% noise}
\end{subfigure}
\caption{Hyperelasticity: state evolution for out-of-training distribution (gating, hidden state, dissipation, power discrepancy from left to right).}
\label{fig:elastic_evolution}
\end{figure}

\fref{fig:elastic_evolution} plots 32 examples of the 9 hidden state components and gates for the 4 cases listed in \tref{tab:rmse}, where ``low gate'' and ``high gate'' use $\lambda_\text{gate} = 10^{-6}$ and $10^{-4}$ respectively while the ``noise'' case increases the additive noise from $\varsigma=10^{-4}$ to $10^{-4}$ for the high gate case.
In all four cases, the power discrepancies \eqref{eq:power_discrepancy} are comparable and evenly distributed around a zero mean.
In all but the high gate case, there is nonzero fictitious dissipation.
In the no-gate case, we see strong transients and then steady internal states, where the hidden states appear to cluster into 4 groups.
The low gate case appears to have 4 corresponding clusters of gate evolutions.
In this case, some gates are off for the entire set of trajectories and the hidden trajectories are similar to those in the ungated case albeit with delayed transients and clustered around zero.
In fact, the low gating minimizes the fictitious dissipation compared to the ungated case (note the scale in \fref{fig:elastic_evolution}c vs. \fref{fig:elastic_evolution}b).
The high gate case is a smooth extension of the low gate case to the limit of all gates off for the duration and no fictitious dissipation.
Clearly with sufficient penalization the expected lower state complexity, and conservative evolution are attained.
More exploration of this behavior will be given in the next demonstration.

We also examined the effect of additive noise on the gating mechanism.
The conservative behavior of the gated ISV-NODE model is robust to noise; its predictions stay conservative up to $\varsigma \approx 10^{-2}$ at which the high-frequency noise triggers flow.
We conjecture at some scale the data appears to be dissipative and/or the noise obscures the conservative nature of the response so that one of the plethora of non-zero flow solutions becomes preferred at the chosen $\lambda_\text{gate} = 10^{-4}$.
Further increasing the penalty will make the zero-flow solution preferable but at some value of $\lambda_\text{gate}$ the accuracy will suffer; this will be explored in the next demonstration.

\begin{table}[]
\centering
\begin{tabular}{|c|r|}
\hline
NN type        & RMSE \\
\hline
no gate        & 0.0688  \\
low gate       & 0.0099  \\
high gate      & 0.0061  \\
\hline
gate+noise     & 0.0097  \\
\hline
\end{tabular}
\caption{Errors on validation data for ISV-NODEs with and without gating.}
\label{tab:rmse}
\end{table}

\subsection{Viscoelastic}

\begin{figure}
\centering
\begin{subfigure}[b]{0.95\textwidth}
\centering
% ././Figures/viscoelastic_gate_accuracy_cycle.pdf ==> fig5a.pdf
\includegraphics[width=0.95\textwidth]{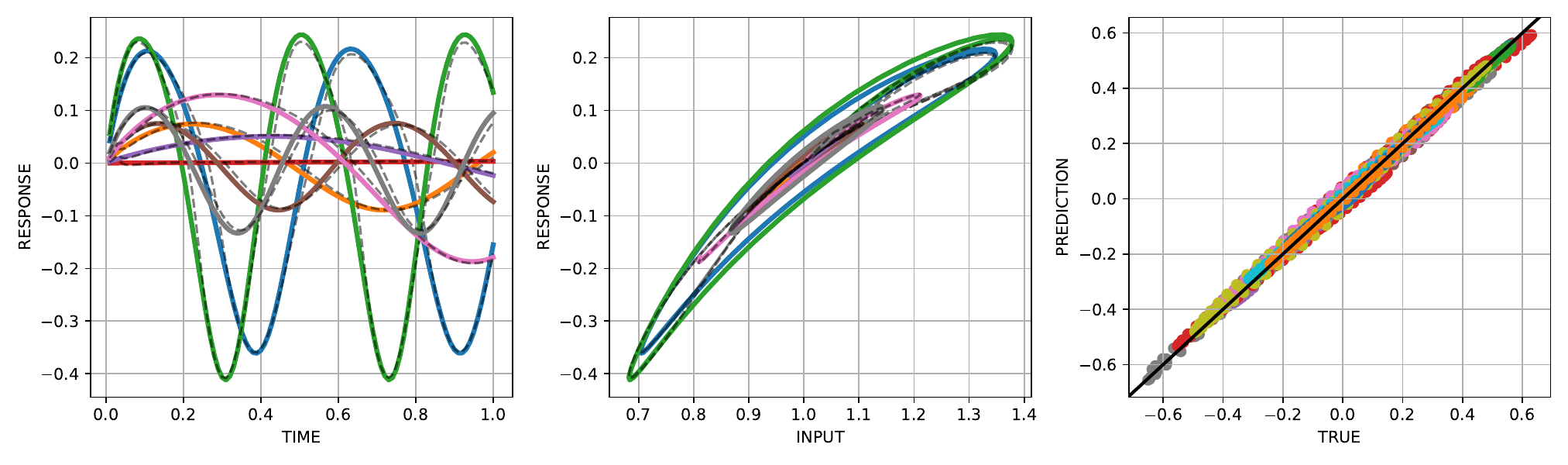}
\caption{accuracy}
\end{subfigure}
\begin{subfigure}[b]{0.95\textwidth}
\centering
% ././Figures/viscoelastic_gate_evolution.pdf ==> fig5b.pdf
\includegraphics[width=0.95\textwidth]{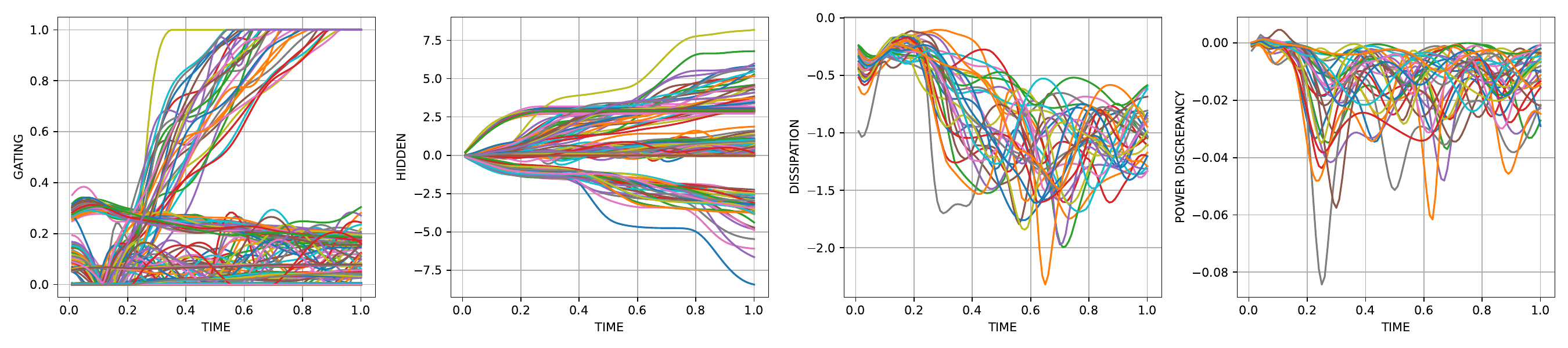}
\caption{evolution}
\end{subfigure}
\caption{Viscoelasticity: (a) accuracy and (b) evolution.
The predictions of stress response (left panel, dashed black) are compared to data (color) versus time in (a) and versus stretch in (b), while (c) plots predicted stress samples versus corresponding true values where samples from the same trajectory have the same color.
}
\label{fig:visco_accuracy}
\end{figure}

\fref{fig:visco_accuracy}a illustrates the accuracy of the nominal gated ISV-NODE on the training data by plotting predicted trajectories and corresponding data trajectories vs time (left panel), stress-stretch cycles (center panel) , and pointwise parity (right panel).
The gated ISV-NODE model is able to well represent the out-of-distribution data with minor amplitude and phase discrepancies despite significant physical dissipation, as evidenced by the stress-stretch hysteresis.
\fref{fig:visco_accuracy}b shows that the gating is quite active.
Some hidden states are turned off as unnecessary, others are on at a constant gating value and still others are amplified over time.
The hidden states evolve at the periodicity of the loading and the dissipation also resembles the sinusoidal loading.
There is also an apparent bias to under-predict power \eqref{eq:power_discrepancy}.

\begin{figure}
\centering
% ././Figures/viscoelastic_Lcurve.pdf ==> fig6a.pdf
\includegraphics[width=0.45\textwidth]{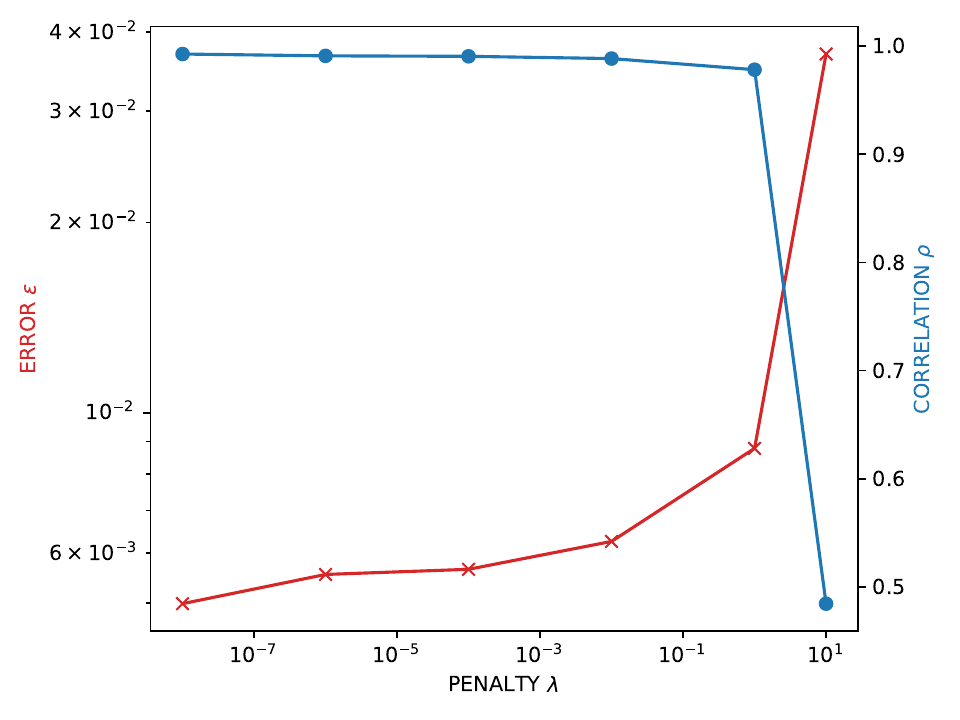}
\caption{Viscoelasticity: accuracy in terms of error (left) and correlation (right) vs penalty $\lambda_\text{gate}$ illustrating the accuracy-complexity tradeoff.}
\label{fig:visco_complexity}
\end{figure}

We explored the sensitivity of the results to the magnitude of the gating penalty $\lambda_\text{gate}$.
\fref{fig:visco_complexity} shows a classical
L-curve \cite{engl1994using} in terms of accuracy versus $\lambda_\text{gate}$.
It is apparent that the ISV-NODE for this data is
relatively insensitive to $\lambda_\text{gate}$ in that the accuracy does not suffer over many orders of magnitude of the penalty.

\subsection{Elastoplasticity}

This last demonstration uses elastoplastic data and a validation dataset with sawtooth trajectories that better illustrate the plastic phases than the sinusoidal used in the two previous demonstrations.

\begin{figure}
\centering
\begin{subfigure}[b]{0.95\textwidth}
\centering
% ././Figures/plastic_nogate_train-accuracy_cycle.pdf ==> fig7a.pdf
\includegraphics[width=0.8\textwidth]{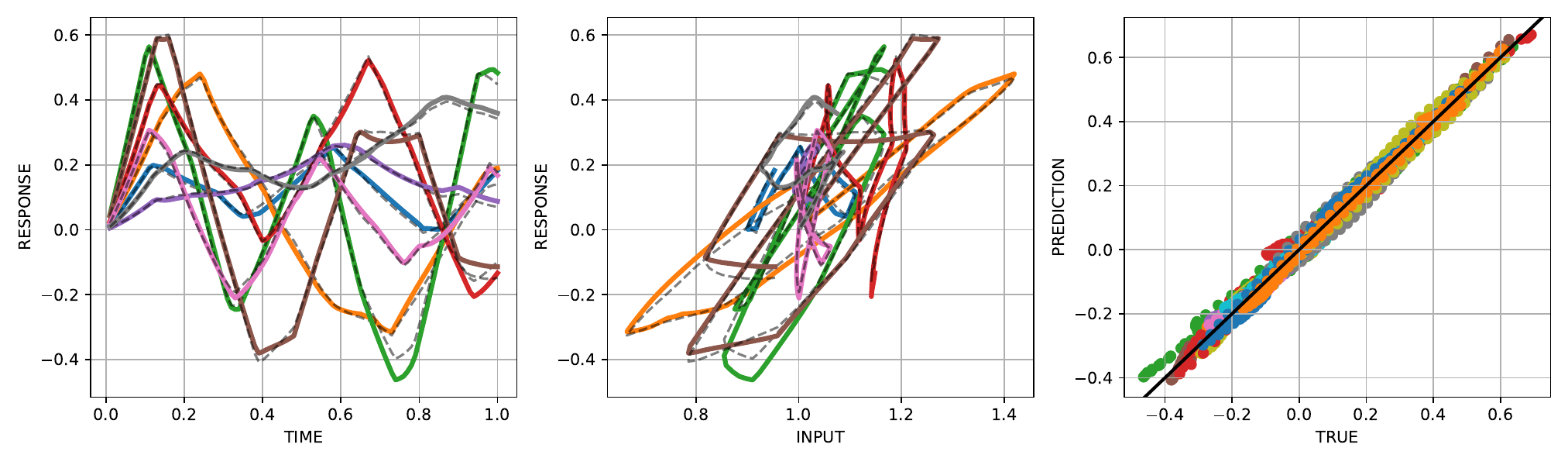}
\caption{no gate}
\end{subfigure}
\begin{subfigure}[b]{0.95\textwidth}
\centering
% ././Figures/plastic_gate_train-accuracy_cycle.pdf ==> fig7b.pdf
\includegraphics[width=0.8\textwidth]{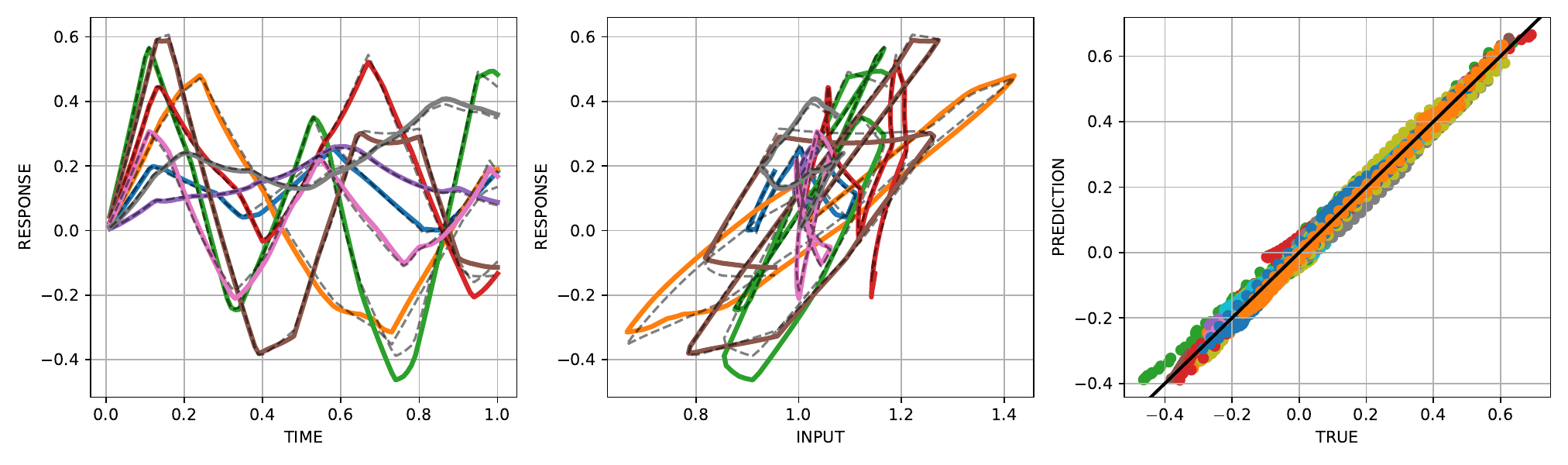}
\caption{gate}
\end{subfigure}
\caption{Elastoplasticity: accuracy (left) without a gate, and (right) with a gate.}
\label{fig:plastic_accuracy}
\end{figure}

\fref{fig:plastic_accuracy} illustrates the accuracy of the non-gated and gated ISV-NODE.
As in the elastic demonstration, they are visibly indistinguishable in terms of the output quantity of interest: stress, as well as quantitatively (validation RMSE: 0.0148  for the nongated ISV-NODE and 0.0145 for the gated model).
Both models capture the elastic and plastic phases well; however, the gated ISV-NODE needs to learn the flow's dependence on the deviatoric stress to be effective.

\begin{figure}
\begin{subfigure}[b]{0.95\textwidth}
\centering
% ././Figures/plastic_nogate_test-evolution.pdf ==> fig8a.pdf
\includegraphics[width=0.9\textwidth]{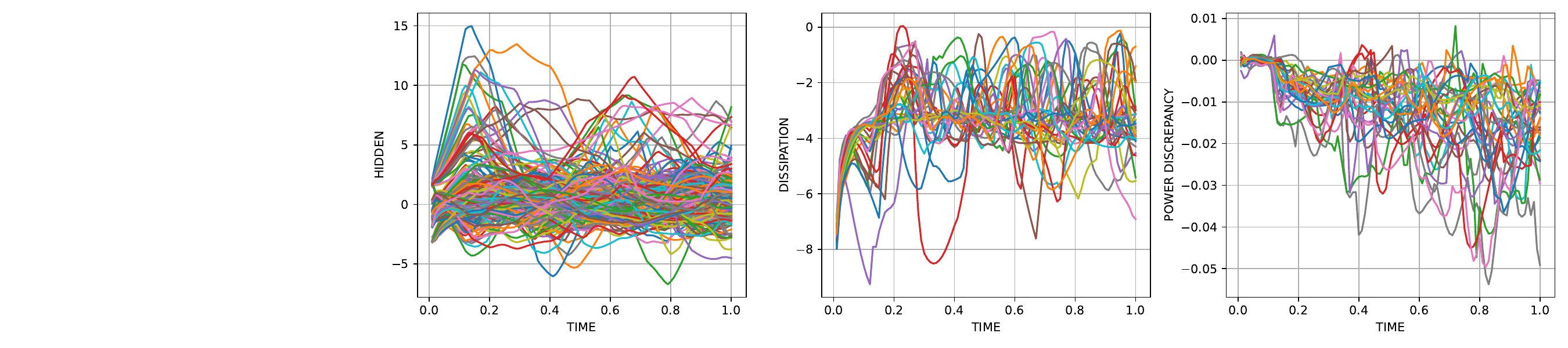}
\caption{no gate}
\end{subfigure}
\begin{subfigure}[b]{0.95\textwidth}
\centering
% ././Figures/plastic_gate_test-evolution.pdf ==> fig8b.pdf
\includegraphics[width=0.9\textwidth]{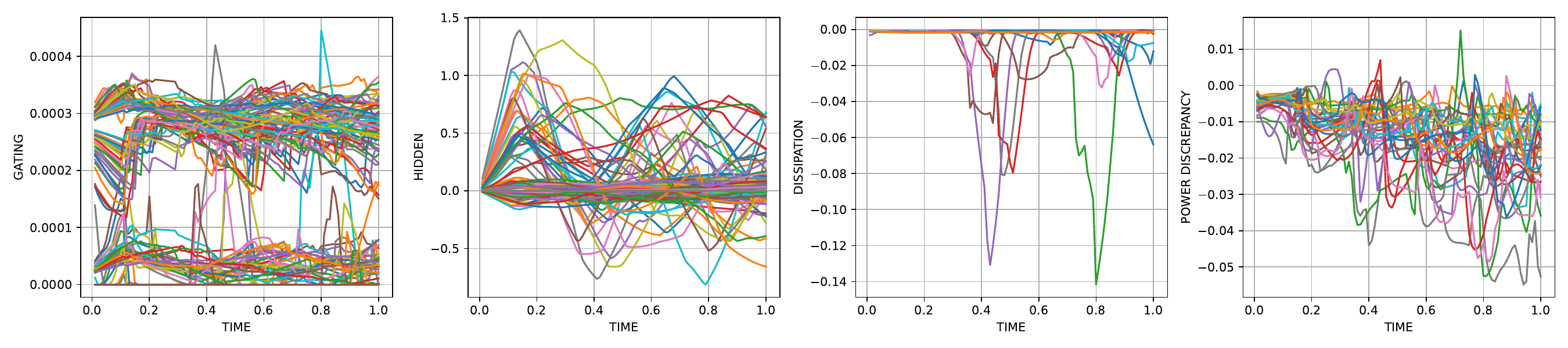}
\caption{gate}
\end{subfigure}
\caption{Elastoplasticity: state evolution (left) without a gate, and (right) with a gate.}
\label{fig:plastic_evolution}
\end{figure}

The trajectories shown in
\fref{fig:plastic_evolution} contrast the evolution of the internal state.
Without gating, the hidden states are continually evolving and the dissipation is always non-zero, which is not the expectation from an elastic-plastic material.
The gated ISV-NODE has the gates actively switching, with some gates staying close to off.
Unlike the ungated ISV-NODE, the gated ISV-NODE there are clear regimes of dissipation and non-dissipation.
As in the other dissipative material (the previous viscoelastic demonstration), there is an apparent bias to under-predict power for both model variants.

\section{Conclusion} \label{sec:conclusion}

The enhancements to the ISV-NODE \cite{jones2022neural} proposed in this work are multifold: (a) a partial ICNN stress potential and (b) an inferred internal state model that uses common latent features to inform novel attention-based gating and drive the flow of internal state.
We demonstrated that this architecture can accurately model dissipative and conservative behavior across an isothermal elastic-viscoelastic-elastoplastic spectrum.
Unlike RNN approaches the internal states and gating mechanism of the ISV-NODE are closer to classical constitutive theory (without being hindered by prescribed internal states) and therefore more interpretable.

A number of alternatives and extensions are apparent.
In this work, we did not pursue a dissipation potential  $\Phi$ (such that $\dot{\internalstate} = \partialb_\conjugateforce \Phi$) beyond preliminary studies.
This variant would need to form $\Phi$ from a partial ICNN of $\conjugateforce$ and $\internalstate$ as in \eref{eq:dissipation_potential}  and then gate this flow to achieve similar results.
It may be possible to formulate a strictly dissipative flow from a potential without the complexity of conjugate forces and the derivatives of derivatives this approach would entail.
Likewise, we did not pursue a bespoke gating based on stress:  $\dot{\internalstate} = \gb(\stress) \fb(\strain,\dot{\strain})$, (or the energy potential) directly since this approach is less attractive due to assumption of prior knowledge of the form of the flow for traditional elastoplastic models.
It may be beneficial to merely expand the inputs with the predicted stress and the predictor-corrector integrator we adopted facilitates this.
Furthermore, the ISV-NODE framework may benefit from inferring the tensorial structure of the internal state as in the multiplicative split \eqref{eq:FeFp}, which is currently assumed to be scalar invariants; however,  no general theory that we are aware of guides this and taking advantage of an invariant based formulation and a differentiable framework is convenient.

Next, we intend to focus on extending the ISV-NODE to the full thermomechanical setting and the attendant thermal-viscous effects.
In addition, we will investigate additional complexity-reducing techniques; the generalizability and efficiency of the ISV-NODE would likely benefit from applying parameter-pruning/sparsification techniques
\cite{fuhg2023extreme,padmanabha2024improving,padmanabha2024condensed} to all the subcomponents.
Learning material symmetry \cite{fuhg2022learning}, phase transitions, and homogenized failure, fracture, and damage also present feasible challenges for extensions of the ISV-NODE framework.

\section*{Acknowledgments}

Sandia National Laboratories is a multimission laboratory managed and operated by National Technology and Engineering Solutions of Sandia, LLC., a wholly owned subsidiary of Honeywell International, Inc., for the U.S.
Department of Energy's National Nuclear Security Administration under contract DE-NA-0003525.
This paper describes objective technical results and analysis.
Any subjective views or opinions that might be expressed in the paper do not necessarily represent the views of the U.S.  Department of Energy or the United States Government.

%%%%%%%%%%%%%%%%%%%%%%%%%%%%%%%%%%%%%%%%%%%%%%%%%%%%%%%%%%%%%%%%%%%%%%%%

%%%%%%%%%%%%%%%%%%%%%%%%%%%%%%%%%%%%%%%%%%%%%%%%%%%%%%%%%%%%%%%%%%%%%%%%

\end{document}